\documentclass[useAMS,usenatbib]{mnras}
\usepackage{deluxetable}

\title[VLA observations of hotspots]{Jansky VLA observations of
  synchrotron emitting optical hotspots of 3C\,227 and 3C\,445 radio galaxies}
\author[M. Orienti et al. ]
  {M. Orienti$^{1}$\thanks{E-mail: orienti@ira.inaf.it},
    G. Migliori$^{1,2}$, G. Brunetti$^1$, H. Nagai$^{3}$,
F. D'Ammando$^{1}$, 
\newauthor K.-H. Mack$^{1}$, M.A. Prieto$^{4}$\\
$^{1}$Istituto di Radioastronomia - INAF, Via P. Gobetti 101, I-40129 Bologna, Italy\\
$^{2}$Dipartimento di Fisica e Astronomia, Universit\`a di Bologna,
Via Gobetti 93/2, I-40129 Bologna, Italy\\
$^{3}$National Astronomical Observatory of Japan, Osawa 2-21-1, Mitaka, Tokyo 181-8588, Japan\\
$^4$Instituto de Astrof\'{i}sica de Canarias, c/ V\'{i}a L\'actea s/n,
E-38205  La Laguna (Tenerife), Spain\\
}
\date{Received \today; accepted ?}

\pagerange{\pageref{firstpage}--\pageref{lastpage}} \pubyear{2002}

\def\LaTeX{L\kern-.36em\raise.3ex\hbox{a}\kern-.15em
    T\kern-.1667em\lower.7ex\hbox{E}\kern-.125emX}

\begin{document}

\label{firstpage}

\maketitle

\begin{abstract}

  We report results on deep Jansky Very Large Array
  A-configuration observations at 22 GHz of the hotspots of the
  radio galaxies 3C\,227 and 3C\,445. Synchrotron emission in the
  optical on scales up to a few kpc was reported for the four
  hotspots. Our VLA observations 
  point out the presence of unresolved regions with upper limit to
  their linear size of about 100 pc. This is the first time that such
  compact components in hotspots have been detected in a mini-sample,
  indicating that they are not a peculiar characteristic of a few
  individual hotspots.   
  The polarization may reach values up to 70 per
  cent in compact compact (about 0.1 kpc scale) regions within the
  hotspot, indicating a   
  highly ordered magnetic field with size up to a hundred parsecs.
  On larger scales, the average polarization of the hotspot
  component is about 
  30$-$45 per cent, suggesting the presence of a significant random
  field component, rather than an ordered magnetic field. This is
  further supported by the displacement between the peaks in polarized
  intensity and in total intensity images that is observed in all the
  four hotspots. The electric vector position angle is not constant,
  but changes arbitrarily in the central part of the hotspot regions,
  whereas it is usually perpendicular to the total intensity contours of the
  outermost edge of the hotspot structure, likely marking the
  large-scale shock front. The misalignment between X-ray and
  radio-to-optical emission suggests that the former is tracing
  the current particle acceleration, whereas the latter
  marks older shock fronts.

\end{abstract}

\begin{keywords}
radio continuum: galaxies - radiation mechanisms: non-thermal -
acceleration of particles - polarization
\end{keywords}

\section{Introduction}

Hotspots are bright and compact regions that are usually
  observed at the edge of 
powerful Fanaroff-Riley type II radio galaxies \citep{fr74}. In the
standard scenario, hotspots represent the working surface of the
supersonic jets produced by an active galactic nucleus (AGN). In this
region both a bow-shock and a reverse shock are present, being the
latter responsible for the acceleration of the particles that radiate
in the hotspot.
Although hotspots mainly radiate
at radio wavelengths, there is a growing evidence that many of them
can emit in infrared/optical band and up to X-rays
\citep[e.g.][]{meise97,hardcastle04,kataoka05,mack09,werner12,isobe17}.
However, there 
seems to be a dichotomy in the radiation mechanism responsible for X-ray
emission between high-power {(L$_{\rm 1.4 GHz} > 10^{25}$ W Hz$^{-1}$)}
and low-power hotspot
\citep{hardcastle04}. X-ray emission in 
high-power hotspots is well explained as synchrotron self-Compton
\citep[SSC,][]{harris94,werner12}, whereas in low-power hotspots an
additional contribution from synchrotron radiation from a highly
energetic electron population is expected to
be present \citep[e.g.][]{hardcastle16}.  \\
\indent Given the anticorrelation between the synchrotron
  radiative lifetime, $\tau$, and the Lorentz 
  factor of the relativistic electrons $\gamma > 10^{7-8}$, 
the detection 
of synchrotron X-ray emission from a hotspot implies 
that particle acceleration is currently taking place owing to the
short radiative lifetime of the electrons involved ($\tau
  \propto$ 1/$\gamma$). Using Very Long
Baseline Array (VLBA) observations, \citet{tingay08}
suggested that X-ray emission in the western hotspot of Pictor\,A may
be produced by synchrotron emission from compact pc-scale transient
regions of enhanced magnetic field. The discovery of flux variability
up to 10 per cent level in the hotspot of Pictor\,A strongly supports
this scenario \citep{hardcastle16}. The timescale of the temporal
variability corresponds to spatial scales much smaller than the
physical size of the hotspot, suggesting that a significant
contribution to the X-ray emission comes from one or more very compact
regions. \\
\indent The detection by the Very Large Telescope (VLT) of kpc-scale diffuse
near-infrared (NIR) and optical synchrotron emission from hotspots provides 
complimentary information on the complexity of particle acceleration
and transport in radio hotspots \citep{mack09}. In these cases a
combination of multiple and intermittent compact acceleration sites and
stochastic (e.g. Fermi-II type) acceleration mechanisms have been
proposed to circumvent the tension between the short lifetime of
the optical synchrotron emitting electrons and the extension of the
emitting region \citep[e.g.][]{prieto02,orienti12}. This idea has been recently
supported by full-polarization Atacama Large Millimeter Array (ALMA)
observations of the hotspot 3C\,445 South where bright and highly
polarized components are enshrouded by diffuse unpolarized emission
\citep{orienti17}.\\ 
In this paper we present the results on new A-configuration Very Large
Array (VLA) 
observations at 22 GHz in full polarization of the hotspots of the
radio galaxies 3C\,227 and 3C\,445 with the aim of understanding how
strong shocks are distributed in the hotspot regions.
These hotspots are part of the low-power (P$_{\rm 1.4\; GHz} < 10^{25}$
W Hz$^{-1}$) hotspot sample selected by \citet{brunetti03}, and target
of VLT K-band observations. NIR emission is detected in each hotspot
\citep{mack09} and is extended on kpc-scale
\citep{orienti12,migliori19}. All the hotspots 
have X-ray emission 
\citep[e.g.][]{hardcastle07, perlman10, orienti12}, 
with the exception of 3C\,445 North which has only a tentative {\it
  Chandra} detection \citep{mingo17}.\\
The high resolution of our VLA
observations allows us to pick up particle acceleration sites down to
scales of few tens parsecs. The availability of polarization
information enables the study of the magnetic field structure on
scales that have never been investigated in hotspots with such details.\\
This paper is organized as follows: in Sections 2 and 3 we present the
observations and data analysis; results are reported in Section 4 and
discussed in Section 5. A brief summary is presented in Section 6.

Throughout this paper, we assume the following cosmology: $H_{0} =
71\; {\rm km/s\, Mpc^{-1}}$, 
$\Omega_{\rm M} = 0.27$ and $\Omega_{\rm \Lambda} = 0.73$,
in a flat Universe. The spectral index $\alpha$
is defined as 
$S {\rm (\nu)} \propto \nu^{- \alpha}$. In all the images North is up
and West is right. 3C\,227 and 3C\,445 are at redshift $z$=0.086
  and $z$=0.055, respectively. At the redshift of the sources, 1 arcsec
  corresponds to 1.62 kpc and 1.07 kpc for 3C\,227 and 3C\,445,
  respectively.\\

\section{Radio observations}

Full-polarization VLA observations in A-configuration for 3C\,227 and
3C\,445 were carried out on 2018 March 5 and May 18, respectively
(project code 18A-087). The observations have
a total band width of 8 GHz divided into 4$\times$2-GHz basebands
centred on 19, 21, 23, and 25 GHz. The target on-source observing time
is about 30 min, for a total observing time of 7 h.
The absolute amplitude scale and the absolute polarization
angle were calibrated using the
flux density scale from \citet{pb13a} and the polarization information
from \citet{pb13b} for the calibrator 3C\,286.
Bandpass was calibrated using the calibrator J0927$+$3902 (4C\,39.25)
and J2253$+$1608 
(3C\,454.3) for 3C\,227 and 3C\,445, respectively.
In addition, the instrumental polarization (D-terms) was calibrated
using the unpolarized sources J0713$+$4349 and J2355$+$4950 for
3C\,227 and 3C\,445, respectively. \\
\indent Calibration and data reduction were done using Common Astronomical
Software Applications (\texttt{CASA}) version 4.7.0. The uncertainty
on the flux density due to amplitude calibration 
errors $\sigma_{\rm cal}$ was estimated by checking the scatter of
  the amplitude gain factors, and turned out to be less than 3 per
cent. This value is in agreement with the errors that are usually
given by \citet{pb13a} from the long term monitoring program.\\
\indent Final images have a resolution
of about 0.08$\times$0.07 arcsec$^2$ and 0.11$\times$0.07 arcsec$^2$
for 3C\,227 and 3C\,445, respectively, and an rms of 6 $\mu$Jy beam$^{-1}$
in Stokes I, Q and U. Images in Stokes Q and U have been
used to produce the 
polarization intensity, polarization angle, fractional polarization
and the associated error images. Blanking on the fractional
polarization image was done by 
clipping the pixels of the input images 
with values below three times the rms measured on the off-source image
plane. The uncertainty on 
the polarization angle is about 4 degrees.\\
\indent For all the sources we also produced images (Stokes IQU),
using the natural
weighting algorithm which are most sensitive to extended emission. The
flux density reported in Table \ref{flux} refers to natural weighting images.\\

\section{{\it Chandra} observations}

X-ray studies of the hotspots of 3C\,227 and 3C\,445 have been
presented in several papers
\citep{hardcastle07,perlman10,orienti12,mingo17,migliori19}. Here, 
we retrieved and analyzed the archival {\it Chandra} observations
with the goal of comparing the X-ray morphology with the structures
observed in total intensity and polarization at 22 GHz. 

The X-ray data analysis was performed with the {\it Chandra}
Interactive Analysis of Observation (CIAO) 4.9 software
\citep{fruscione06} using the 
calibration files CALDB version 4.7.7. We ran the
\texttt{chandra$\_$repro} reprocessing script, that performs all the 
standard analysis steps.   

We checked and, when necessary, filtered the data for the time intervals
of background flares. We adjusted the astrometry of the X-ray images
by shifting the X-ray centroid of the cores to the radio positions and
verified the correct alignment by comparison with the positions of
background sources with infrared and optical catalogs \citep[2MASS
  Point Source Catalog and USNO B1.0,
][]{Skrutskie06}. In this way, we
estimated $<$0.15 arcsec pointing accuracy. 

3C\,227 was observed twice in January 2006 (ObsIDs 7265 and 6842) in
very faint (VF) mode for $\sim$53 ksec in total and the two
observations were merged together and rebinned to a 
pixel size of 0.123 arcsec.

We selected the $\sim$46 ksec observation of 3C\,445 performed by {\it
  Chandra} in faint (F) mode in 2007 (ObsID 7869)\footnote{In two
  other epochs of observations of 3C\,445 in 2009 and 2011, the 
pointing settings were not suitable for our study. On these
occasions the use of the gratings reduced the photon counts, and the
hotspots were $>$4 arcmin offset from the observation axis and
thus severely affected by distortions of the point spread function
(PSF).}.
Given the higher number of counts, the X-ray image of
3C\,445 could be rebinned to one eighth of the native pixel size
(0.0615 arcsec) in
order to have a resolution roughly matching that of the VLA data. 

The western hotspot of 3C\,227 and the southern hotspot of 3C\,445 were placed near the aim point, on the S3 chip of the ACIS-S array. Because of the angular extension of the two radio galaxies, the hotspots of the counter jets fall on a different chip and they are only weak or marginally detected in X-rays \citep[see also][]{hardcastle07,perlman10,mingo17}.\\

\begin{table*}
\caption{Total intensity and polarization properties of the
  hotspots. Column 1: hotspot name; Column 2: hotspot component;
  Column 3: total intensity flux density at 22 GHz; Columns 4 and 5:
  polarized flux density and fractional polarization at 22 GHz, respectively.}
\begin{center}
\begin{tabular}{llccc}
  \hline
  Hotspot&Comp.& I & P & m \\
         &     &mJy&mJy& \% \\
  \hline
  3C\,227 West & E & 27.42$\pm$0.84 & 11.56$\pm$0.39 & 42$\pm$1 \\
               & N &  3.51$\pm$0.11 &  1.47$\pm$0.06 & 42$\pm$1 \\
               & S &  3.80$\pm$0.12 &  1.85$\pm$0.06 & 48$\pm$1 \\
               & W &  6.44$\pm$0.25 &  3.55$\pm$0.18 & 55$\pm$3 \\
  3C\,227 East & HS&  8.15$\pm$0.28 &  2.79$\pm$0.15 & 34$\pm$2 \\
               & K &  0.59$\pm$0.04 &  0.27$\pm$0.04 & 45$\pm$8 \\
               & N &  3.40$\pm$0.16 &  1.76$\pm$0.14 & 52$\pm$4 \\
               & S &  9.33$\pm$0.35 &  5.31$\pm$0.27 & 57$\pm$3 \\
               & V & 14.59$\pm$0.51 & 10.58$\pm$0.41 & 72$\pm$2 \\ 
3C\,445 North &  Tot   &16.67$\pm$0.50 & 7.42$\pm$0.24 & 44$\pm$6     \\
3C\,445 South & Tot    &38.10$\pm$1.15 & 12.24$\pm$0.40& 32$\pm$1\\
              & E      &14.24$\pm$0.43 &  5.18$\pm$0.16& 36$\pm$1\\
              & W      & 2.92$\pm$0.09 &  0.79$\pm$0.04& 27$\pm$1\\
              & S      & 0.90$\pm$0.03 &  0.37$\pm$0.02& 41$\pm$2\\
\hline
\end{tabular}
\end{center}
\label{flux}
\end{table*}

\begin{figure*}
  \begin{center}
\includegraphics{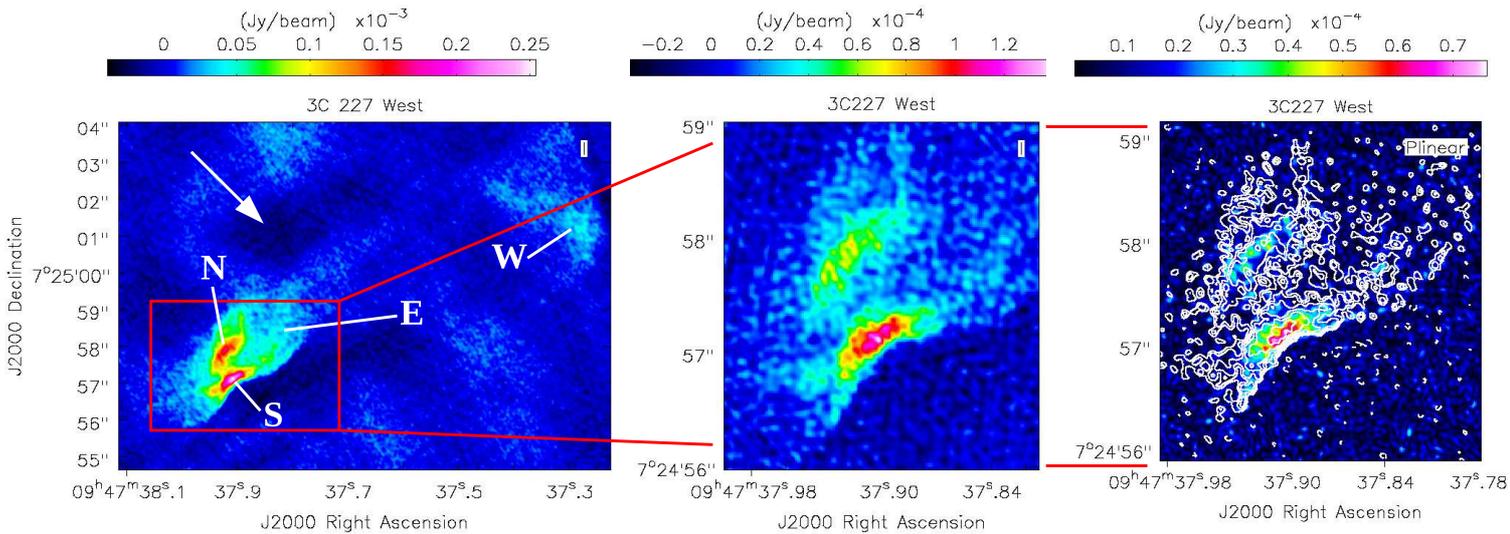}    
\vspace{8.5cm}
\caption{{\it Left panel}: Total intensity VLA image
  at 22 GHz of the hotspot complex 3C\,227 West obtained with natural
  weighting. The arrow indicates the direction of the jet. {\it
    Central panel}: Zoom of the Eastern region of 
  the hotspot 3C\,227 West in total intensity obtained using Briggs
  weighting. {\it Right panel}:   
  Zoom of the Eastern region of the hotspot 3C\,227 West in
  total intensity 
  (contours) overlaid with polarized intensity image. The first
  contour is 18 $\mu$Jy beam$^{-1}$ and corresponds to three times the
off-source noise level measured on the image plane. Contours increase
by a factor of $\sqrt{2}$. The colour scale
is shown by the wedge at the top of each image.}
\label{3C227W_FULL}
  \end{center}
\end{figure*}

\begin{figure}
  \begin{center}
\includegraphics{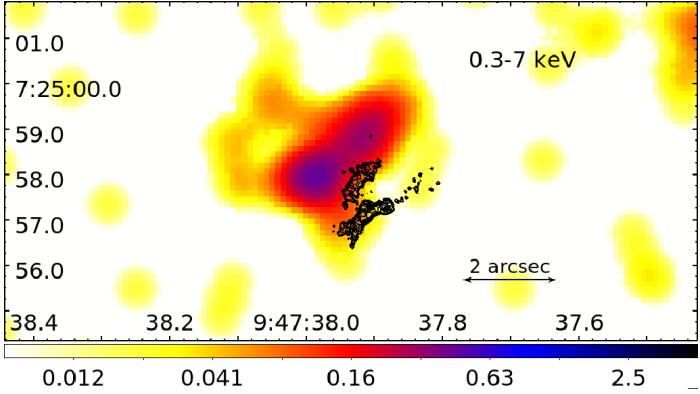}   
\vspace{5.5cm}
\caption{0.3$-$7 keV {\it Chandra} image of the eastern component of
  3C\,227 West (colour-scale) overlaid with 22-GHz VLA image. The first
contour is 24 $\mu$Jy and contours increase
by a factor of $\sqrt{2}$. The colour scale in counts/pix
is shown by the wedge at the bottom of the image.}
  \label{3c227w-chandra}
  \end{center}
  \end{figure}

\begin{figure*}
\begin{center}  
\includegraphics{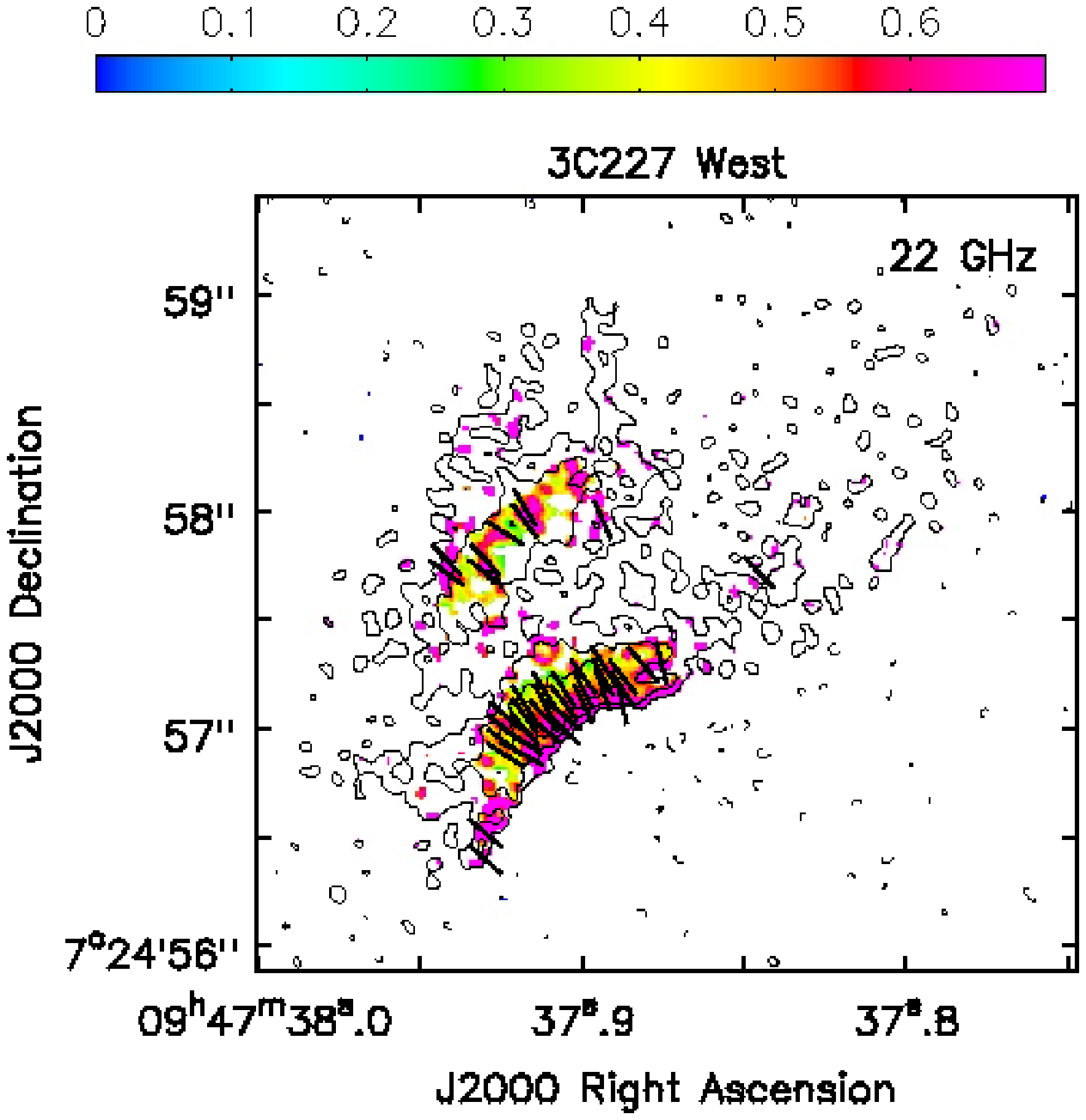}
\includegraphics{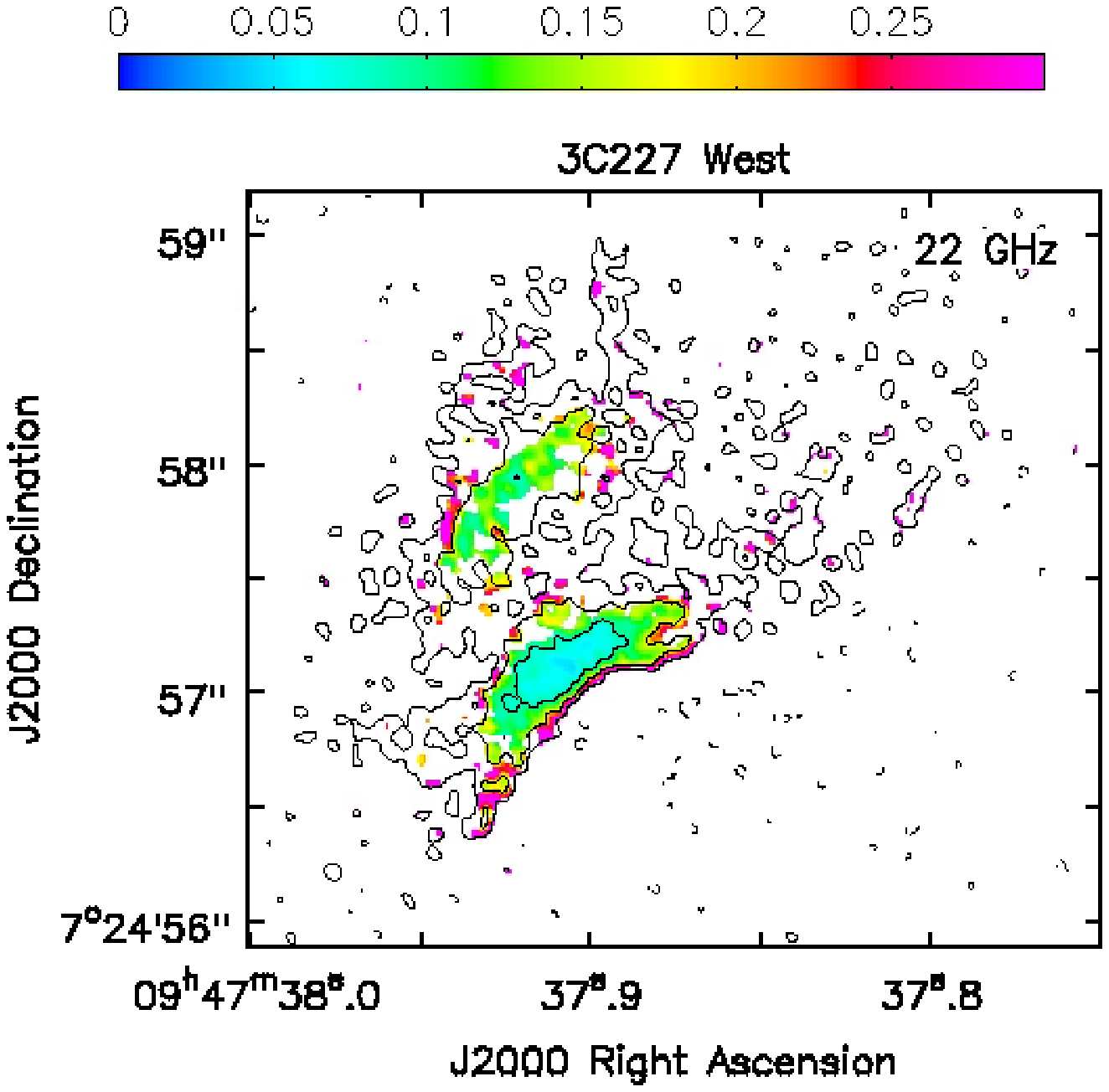}
\vspace{8.5cm}
\caption{Fractional polarization
 image ({\it left}) and fractional
 polarization error image ({\it right}) of the hotspot 3C\,227 West.
  The first
contour is 18 $\mu$Jy beam$^{-1}$ and corresponds 
to three times the off-source noise level measured on the image
plane. Contours increase by a factor of 2. The colour scale
is shown by the wedge at the top of each image. Vectors represent the
electric vector position angle.}
\label{3C227W_POLLA}
\end{center}
\end{figure*}

\begin{figure*}
  \begin{center}
\includegraphics{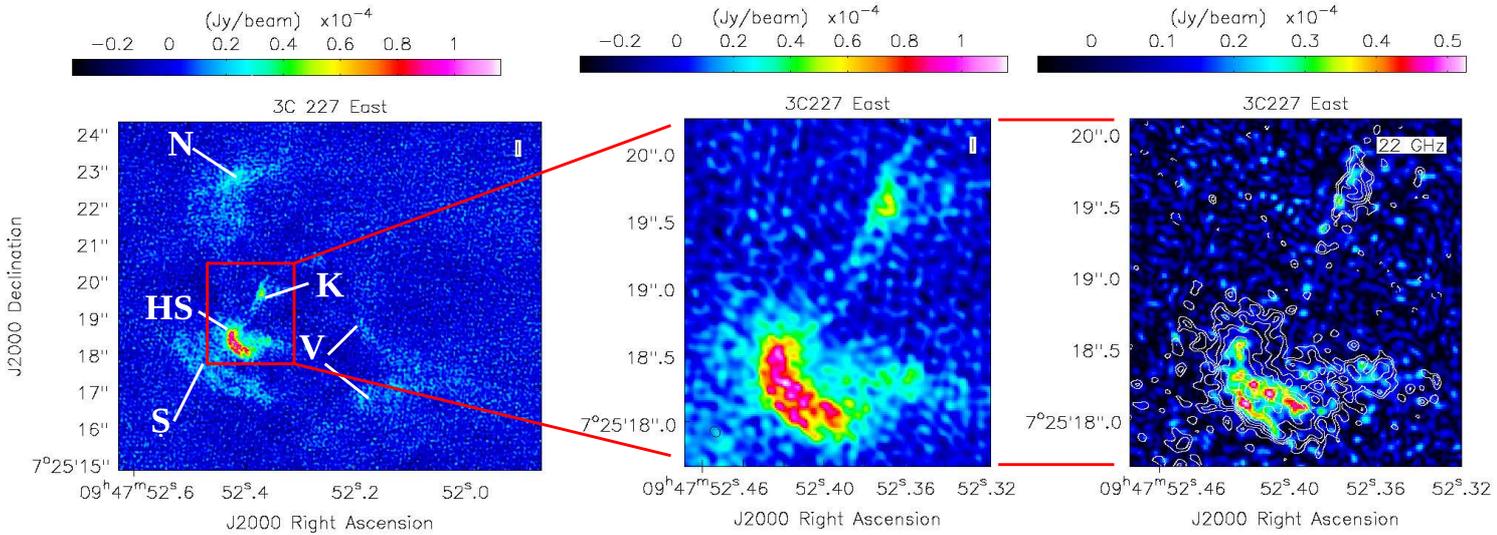}
\vspace{7.5cm}
\caption{{\it Left panel}: Total intensity VLA image
  at 22 GHz of the hotspot complex 3C\,227 East obtained with natural
  weighting. The arrow indicates the direction of the jet. {\it Central panel}: Zoom of the main hotspot region of
  3C\,227 East in total intensity obtained using Briggs
  weighting. {\it Right panel}:   
  Zoom of the main hotspot region of 3C\,227 East in
  total intensity 
  (contours) overlaid with polarized intensity image. The first
  contour is 18 $\mu$Jy beam$^{-1}$ and corresponds to three times the
off-source noise level measured on the image plane. Contours increase
by a factor of $\sqrt{2}$. The colour scale
is shown by the wedge at the top of each image.}
\label{3C227E_FULL}
\end{center}
\end{figure*}

\begin{figure*}
  \begin{center}
\includegraphics{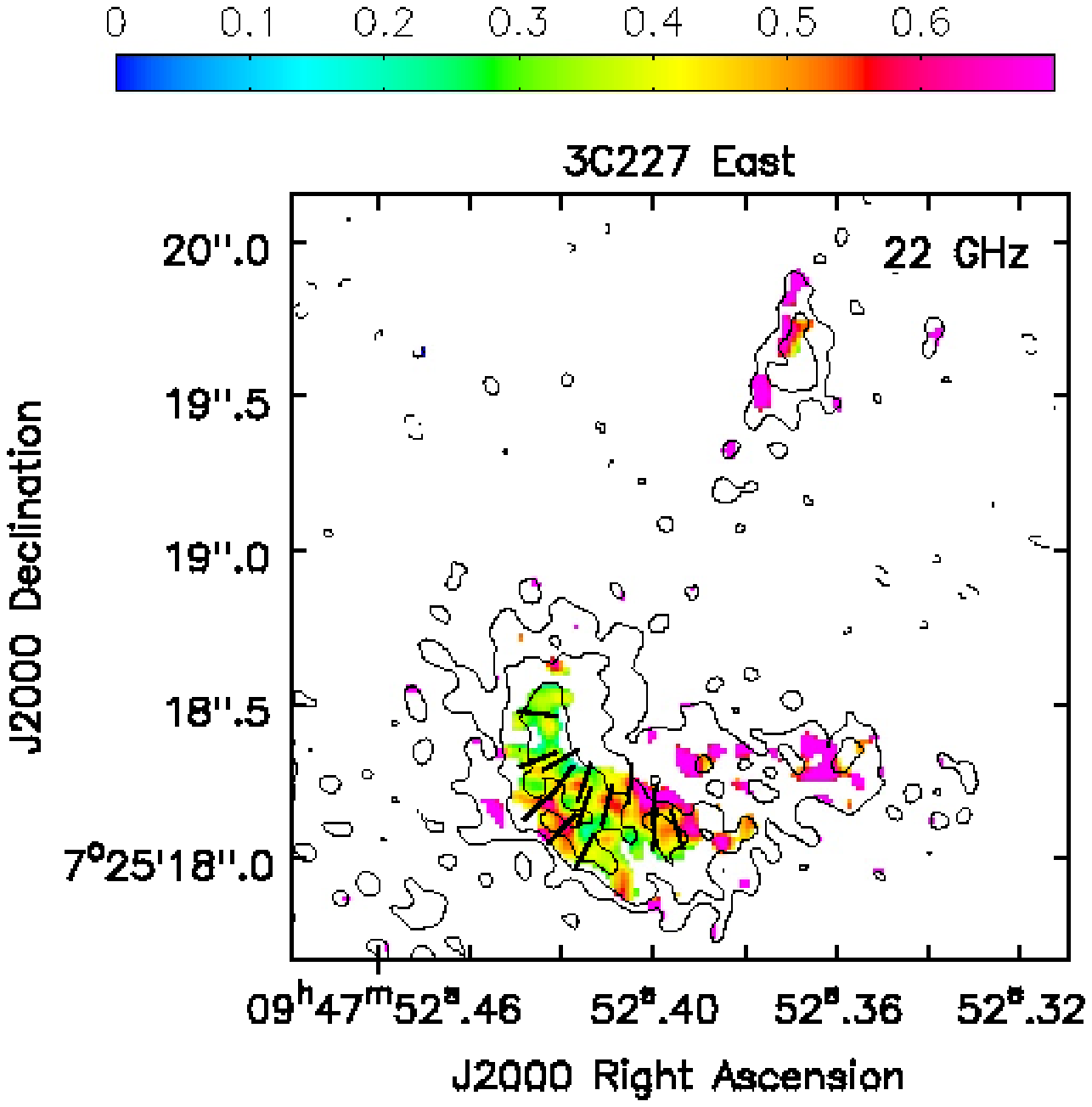}
\includegraphics{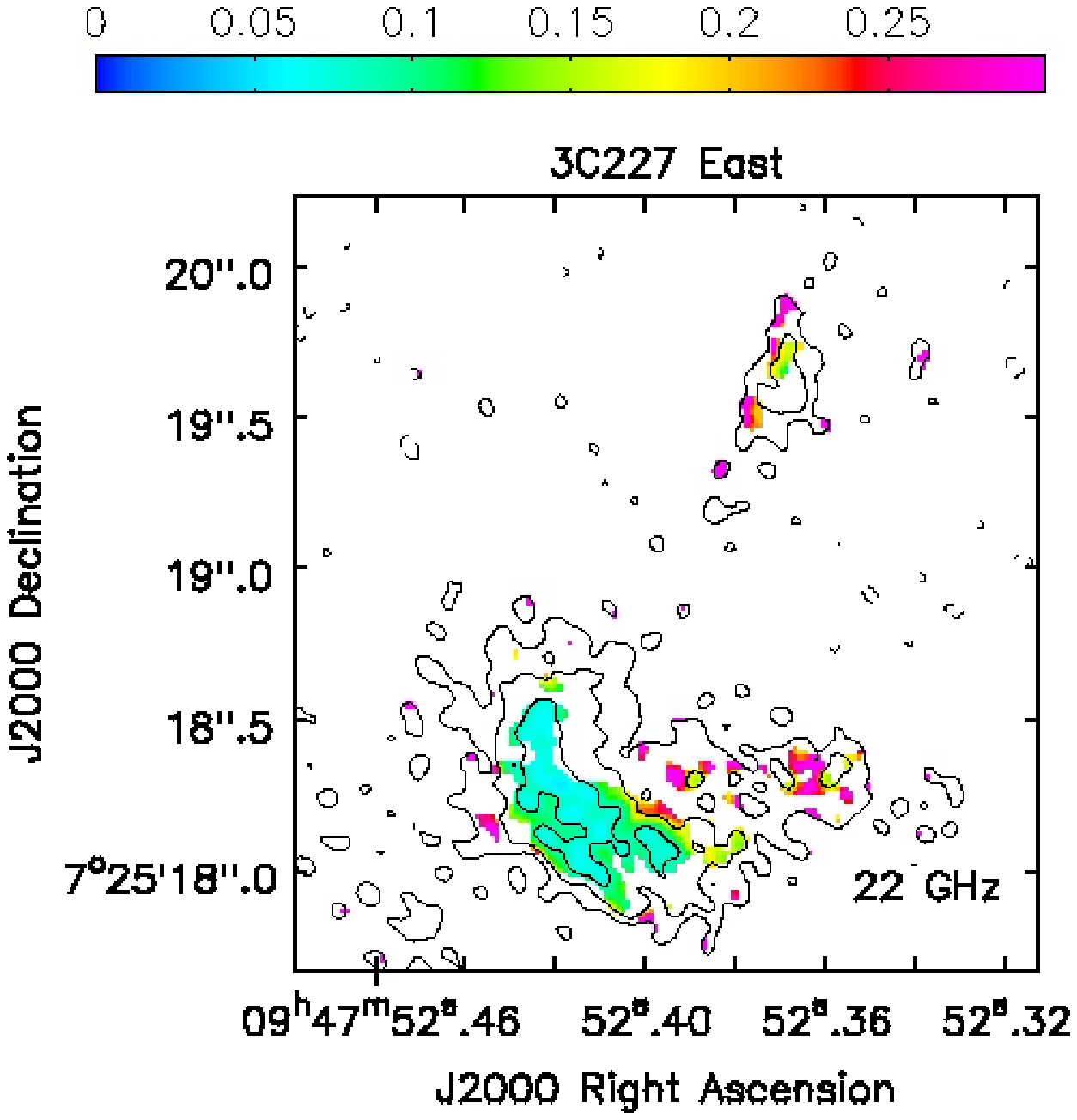}    
\vspace{8.5cm}
\caption{Fractional polarization
 image ({\it left}) and fractional
 polarization error image ({\it right}) of the hotspot 3C\,227
 East.
  The first
contour is 18 $\mu$Jy beam$^{-1}$ and corresponds 
to three times the off-source noise level measured on the image
plane. Contours increase by a factor of 2. The colour scale
is shown by the wedge at the top of each image. Vectors represent the
electric vector position angle.}
\label{3C227E_POLLA}
\end{center}
\end{figure*}

\begin{figure}
  \begin{center}
\includegraphics{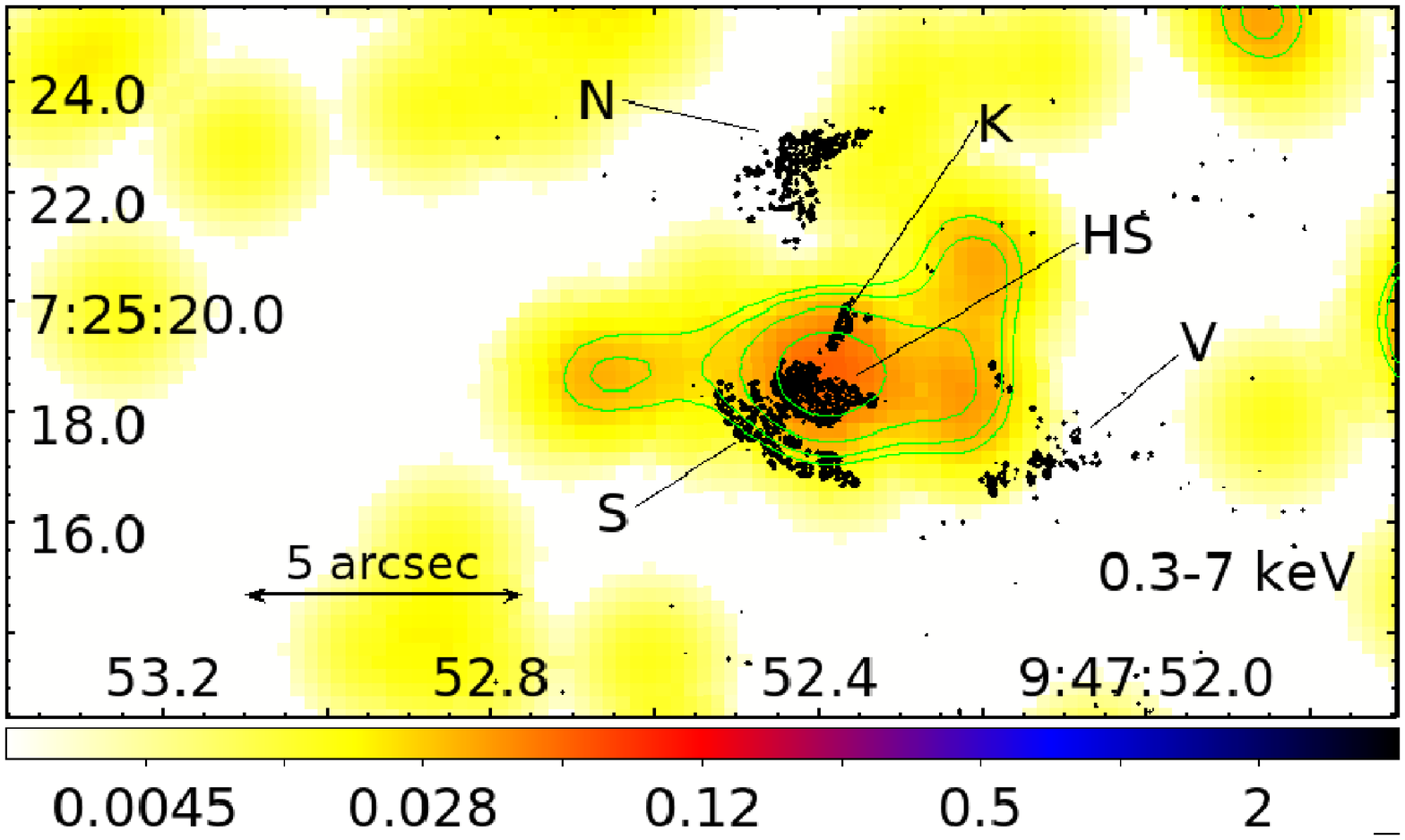} 
\vspace{5cm}
\caption{0.3$-$7 keV {\it Chandra} image of
  3C\,227 East (colour-scale) overlaid on 22-GHz VLA image. The image
  is half the native {\it Chandra} pixel size and has been smoothed
  with a Gaussian filter with $\sigma=7$. The logarithmic X-ray contours (in
  green) start at three times the rms noise (0.007 cts). The first
contour is 18 $\mu$Jy and contours increase
by a factor of $\sqrt{2}$. The colour scale
is shown by the wedge at the bottom of the image.}
  \label{3c227e-chandra}
  \end{center}
  \end{figure}

\section{Results}

Significant radio emission at 22 GHz is observed in all four
hotspots targeted in the study. 
The total intensity flux density and the polarized flux density are
extracted from the same region by selecting polygonal areas on the
image plane. The
values are reported in Table \ref{flux}. We also derive the peak flux
density and the upper limit on the 
deconvolved angular size of the unresolved components by performing a Gaussian
fit to the source components on the image plane using the task \texttt{imfit}
in \texttt{CASA}.      
The error on the total intensity and polarized flux density,
$\sigma_{S}$, is given by $\sigma_{S} = \sqrt{ \sigma_{\rm cal}^2 +\sigma_{\rm rms}^2}$, where $\sigma_{\rm rms}={\rm rms}*\sqrt{N}$ and $N$ is the number of
independent beam areas 
through the component region, while $\sigma_{\rm cal}$ is about
three per cent (see Section 2). The error on the position of peak of unresolved
components is $\Delta \theta = \sqrt{ \sigma_{\rm fit}^2 + \sigma_{\rm
    p}^2}$, where $\sigma_{\rm fit}$ is the error of the
Gaussian fit, whereas 
$\sigma_{\rm p} = {\rm beam size}/{\rm SNR}$, where
SNR is the signal-to-noise ratio and corresponds to S$_{\rm p}$/rms
where S$_{p}$ is the peak flux density of the component and rms corresponds to
1$\sigma$ noise level. For unresolved components with high SNR
$\sigma_{\rm p}$ may be very small. In this case we 
assume a conservative error on the component position of 1/10 of the beam size
\citep[e.g.][]{polatidis03}.\\   
\indent Errors on the polarization parameters
were computed following \citet{fanti01}.
Errors on
the fractional polarization range from about 1 per cent in the
brightest regions up to 40 per cent in the ridges and in  the extended
structures with low SNR.\\ 
In all the hotspot complexes we could
detect hints of extended emission that is present in images at lower
frequencies \citep{black92,leahy97,orienti12}, and that is mainly
resolved out in our observations (largest angular size $\leq$ 2.4
arcsec). Significant polarization is observed
in all the hotspot regions reaching values up to 70 per cent.\\
A description of each hotspot is presented in the following sections.\\

\subsection{3C\,227 West}

3C\,227 West is
a double hotspot with the eastern and western components separated by
about 10 arcsec \citep[e.g.][]{black92}. 
The western component has a size of about 3 arcsec ($\sim$ 4.8 kpc), and is
detected only in 
natural weighting images (Fig. \ref{3C227W_FULL}), 
while it is resolved  
out in full-resolution images.
The high angular resolution of our VLA observations could resolve the
structure of the eastern component for the first time in two
arc-shaped structures elongated in the SE-NW direction, and separated 
by about 0.75 arcsec (1.2 kpc), with the northern component in agreement
with the edge of the X-ray emission (Fig. \ref{3c227w-chandra}). The
elongation of these components is roughly transverse to the direction
of the jet, as it is deduced by low-resolution images \citep[e.g.][]{mack09}.
VLT observations suggest a similar double arc-shaped structure of the
NIR emission, but the lower angular resolution prevents us from a more
 detailed imaging in the NIR/optical regime \citep{migliori19}. The
southern arc, S component in Fig. \ref{3C227W_FULL},
is the brightest one and accounts for $\sim$20 per cent of
the flux density of the hotspot region. The polarization and the total
intensity emission have a similar structure. In the hotspot components
N and S there
are unresolved polarized regions with 
size between 75 and 100 mas, corresponding to about 120$-$160
pc, and a flux density of about 35$-$150 $\mu$Jy. There seems to be a
displacement between the position of the compact components observed
in total intensity and those observed in polarized emission, which
accounts for about 40$\pm$10 mas (65$\pm$7 pc) 
between the total intensity peak and the polarization peak of the
brightest 
unresolved
sub-component of the southern arc (Fig. \ref{3C227W_FULL}, right
panel). 
The polarized emission is patchy and the observed fractional polarization
ranges between 15 and 60 per cent 
(Fig. \ref{3C227W_POLLA}).  
Electric vector position angle (EVPA) is not constant across the
hotspot region, ranging between 
15$^{\circ}$ and 60$^{\circ}$ in component S, and between 30$^{\circ}$
and 60$^{\circ}$ in component N.

\subsection{3C\,227 East}

This hotspot region has a complex structure. The main hotspot
component is labeled HS in Fig. \ref{3C227E_FULL}, is about 0.8 arcsec 
($\sim$1.3 kpc)
in size, and accounts for 30 per cent of the emission measured on the
entire hotspot complex.
Component HS is resolved in several
polarized sub-components (Fig. \ref{3C227E_FULL}, central and right panels).
The upper limit to their size is about
70$-$100 mas, corresponding to about 110$-$160 pc, and the flux density is
about 20$-$50 $\mu$Jy. Although the polarization
and total intensity emission have a similar structure, there seems to
be a displacement
between the peaks of the total intensity emission and the peaks of polarized
emission which accounts for 40$\pm$10 mas (65$\pm$7 pc) for the brightest and
unresolved components, similarly to 3C\,227 West. The average
observed fractional polarization for component HS is about 
34 per cent,
but it reaches values as high as 50 per cent in the central region
(Fig. \ref{3C227E_POLLA}). EVPA changes across component HS, ranging
between 15$^{\circ}$ and $-$70$^{\circ}$ in the highly polarized
structure. 
A bright component is observed at about 1.5
arcsec ($\sim$2.4 kpc) to the North-West 
of component HS, and is labeled K in Fig. \ref{3C227E_FULL}. Component
K may trace the location where the jet bends before reaching
component HS.
The X-ray emission is mainly observed between component HS and K. However,
the low signal-to-noise ratio precludes us from determining the precise
X-ray brightness distribution (Fig. \ref{3c227e-chandra}).\\
In addition to these bright components, three
filaments are clearly visible in our radio images, in agreement with
what was found at 3.6\,cm by \citet{black92}.
A cusp-like structure, labeled V in Fig. \ref{3C227E_FULL}, is
present to the West and seems to mark the edge of the X-ray emission
(Fig. \ref{3c227e-chandra}). A deeper X-ray observation is crucial for confirming this positional correspondence. These extended structures are characterized
by very high level of polarization (Table 
\ref{flux}). This  
may be an indication that the total intensity structure is
extended on scales larger than the largest angular size recoverable by
our observations ($\sim$ 2.4 arcsec) as also suggested by images at
3.6\,cm presented 
in \citet{black92}, while the polarized structures are more
compact.

\subsection{3C\,445 North}

\begin{figure*}
\begin{center}
\includegraphics{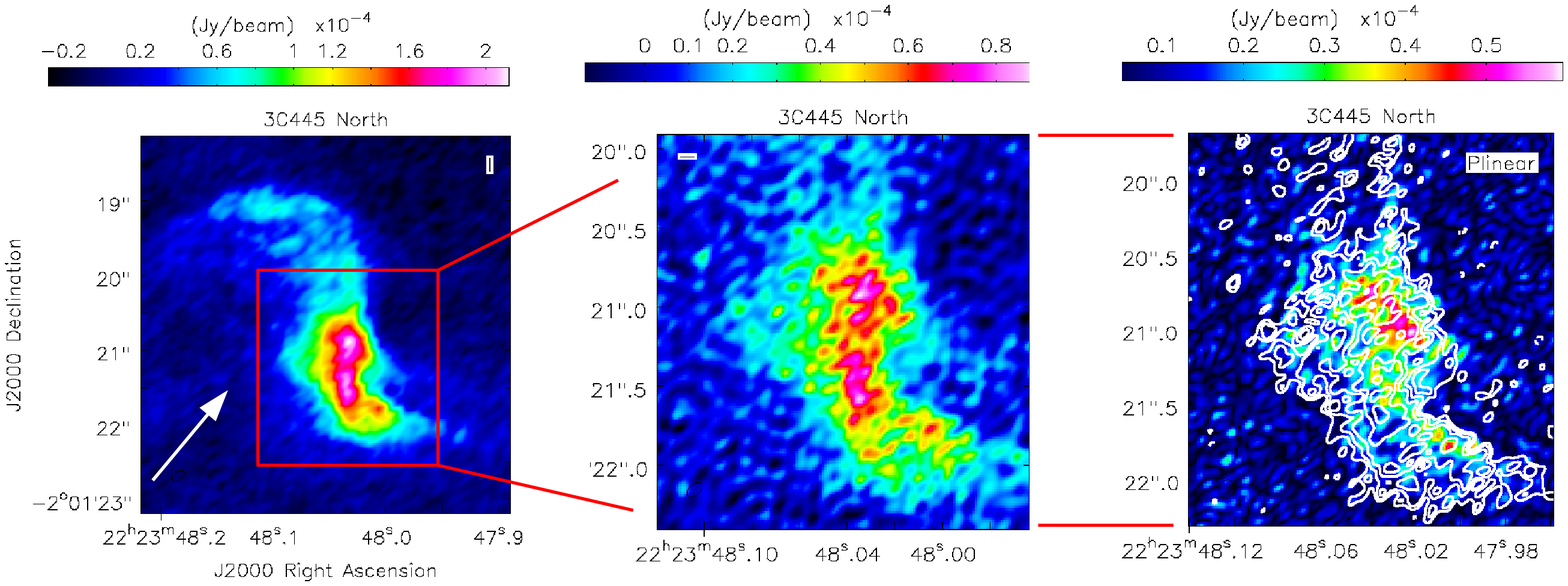}
\vspace{8.0cm}
\caption{{\it Left panel}: Total intensity VLA image
  at 22 GHz of the hotspot complex 3C\,445 North obtained with natural
  weighting. The arrow indicates the direction of the jet. {\it
    Central panel}: Zoom of the central region of the 
  hotspot 3C\,445 North in total intensity obtained using Briggs
  weighting. {\it Right panel}:   
  Zoom of the central region of the hotspot 3C\,445 North in total intensity
  (contours) overlaid with polarized intensity image. The first
  contour is 18 $\mu$Jy beam$^{-1}$ and corresponds to three times the
off-source noise level measured on the image plane. Contours increase
by a factor of $\sqrt{2}$. The colour scale
is shown by the wedge at the top of each image. }
\label{3C445N_FULL}
\end{center}
\end{figure*}

\begin{figure*}
\begin{center}  
\includegraphics{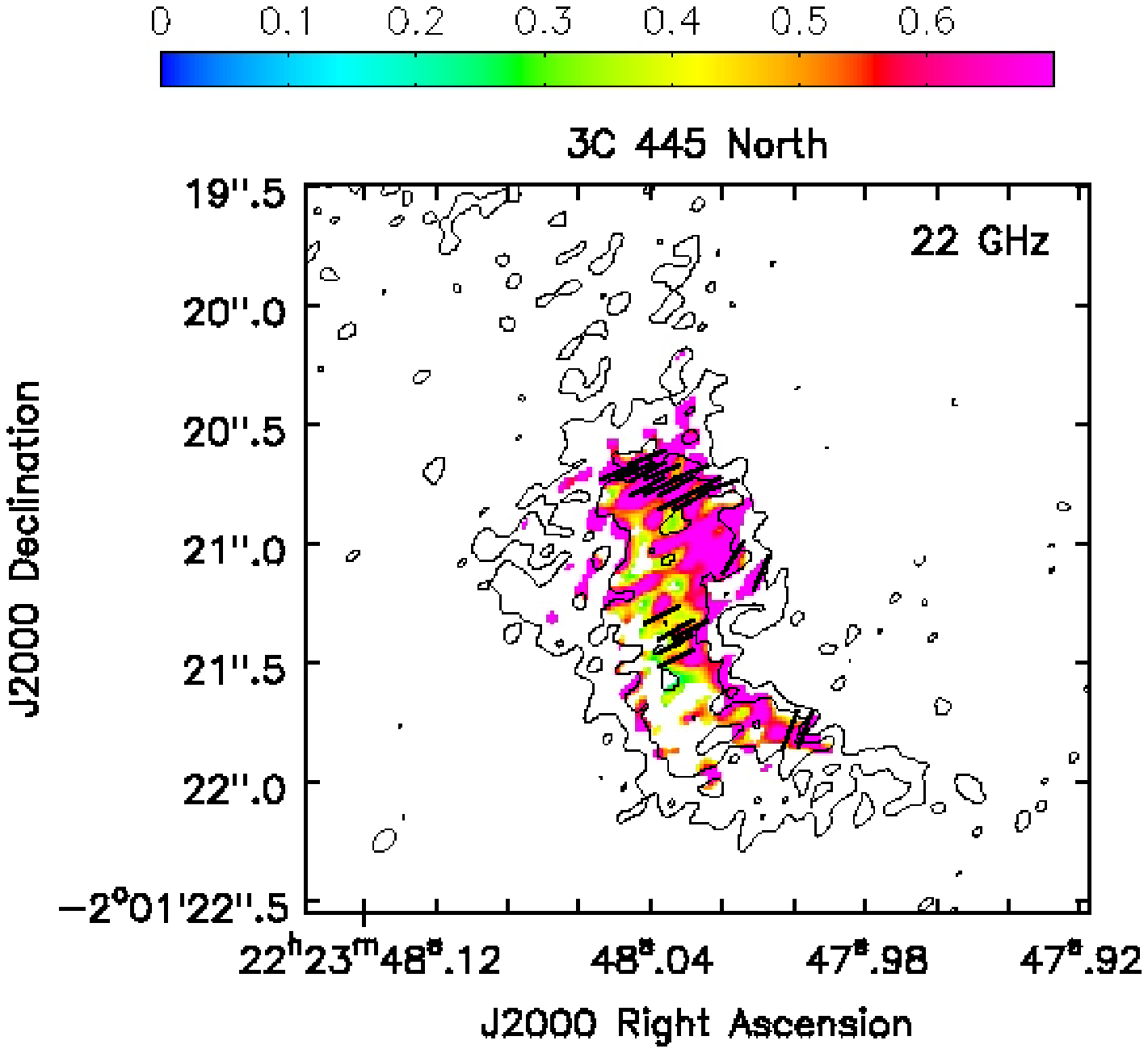}
\includegraphics{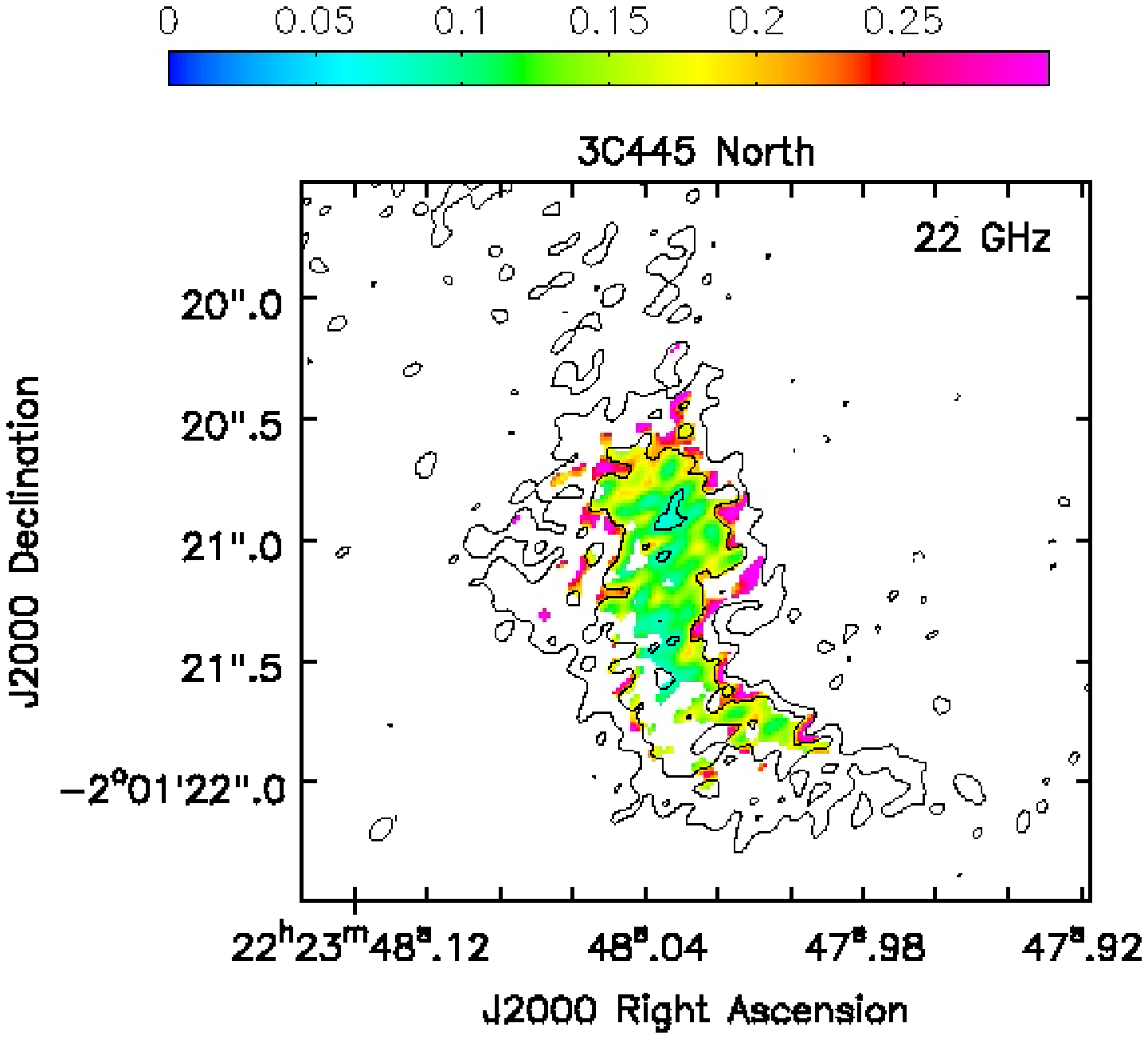} 
\vspace{7.5cm}
\caption{Fractional polarization
 image ({\it left}) and fractional
 polarization error image ({\it right}) of the hotspot 3C\,445 North.
 The first
contour is 18 $\mu$Jy beam$^{-1}$ and corresponds 
to three times the off-source noise level measured on the image
plane. Contours increase by a factor of 2. The colour scale
is shown by the wedge at the top of each image. Vectors represent the
electric vector position angle.}
\label{3C445N_POLLA}
\end{center}
\end{figure*}

This hotspot is about 4 arcsec in size and has an S-shaped structure
elongated in the NE-SW direction, roughly transverse to the jet
direction as derived from the low-resolution images presented in
\citet{mack09}. The radio emission is dominated by
the central region which is resolved in several sub-components
(Fig. \ref{3C445N_FULL}). 
When observed with full-resolution, both the total intensity and the polarized
emission are patchy with several unresolved sub-components enshrouded
by diffuse emission. The 
upper limit to the angular size of these sub-components is about
70$-$120 mas, corresponding 
to a linear size of about 
75$-$130 pc, and
a flux density between 20 and 30 $\mu$Jy.
The total intensity 
structure is elongated in NS direction, while the polarized emission
seems to form two roughly parallel structures oriented in the NE-SW
direction, with the northern component being the brighter
(Fig. \ref{3C445N_FULL}) and accounting for roughly 10 per cent of the
polarization of the entire hotspot.
The average observed fractional
polarization of the hotspot complex is about 44 per cent, reaching
values as high as 60 per cent in the central region.
The EVPA changes across the hotspot region ranging from -32$^{\circ}$
to -85$^{\circ}$, being locally roughly transverse to the elongation
of the polarized structure and to the total intensity contours of the
northern edge of the S-shaped structure (Fig. \ref{3C445N_POLLA}).\\

\subsection{3C\,445 South}

\begin{figure}
  \begin{center}
\includegraphics{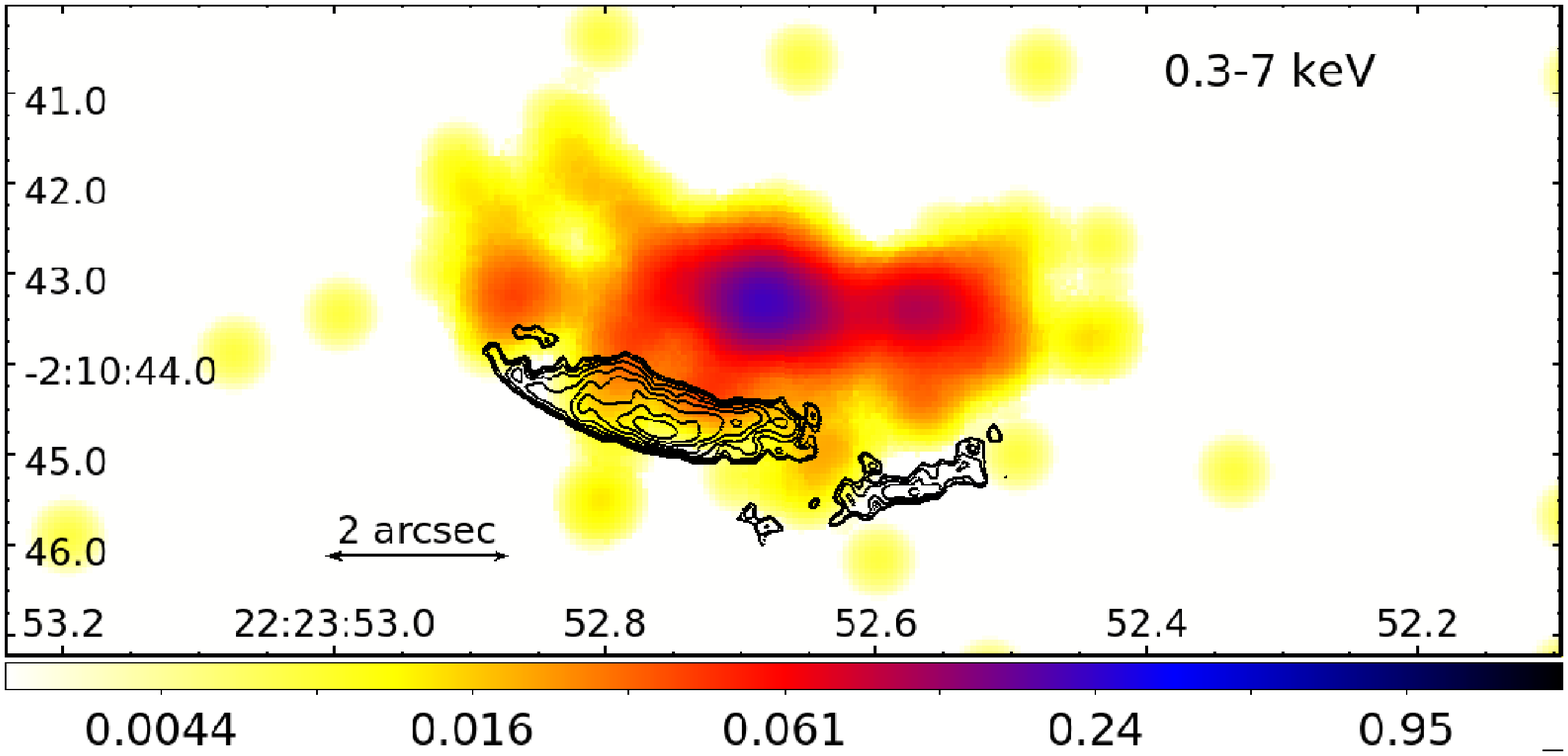}  
\vspace{6cm}
\caption{0.3$-$7 keV {\it Chandra} image of 3C\,445 South (colour-scale)
  overlaid with 22-GHz VLA image. The first 
contour is 18 $\mu$Jy and contours increase
by a factor of $\sqrt{2}$. The colour scale
is shown by the wedge at the bottom of the image.}
\label{3c445s-chandra}
  \end{center}
  \end{figure}

\begin{figure*}
  \begin{center}
\includegraphics{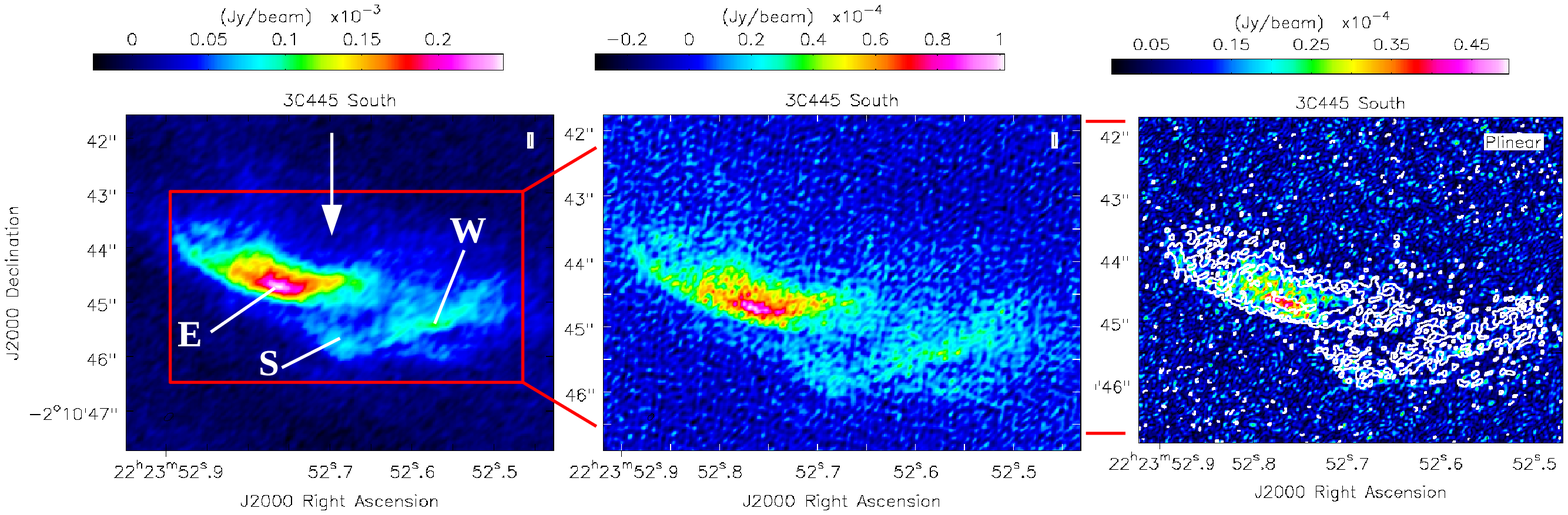}
\vspace{7.5cm}
\caption{{\it Left panel}: Total intensity VLA image
  at 22 GHz of the hotspot complex 3C\,445 South obtained with natural
  weighting. The arrow indicates the direction of the jet. {\it
    Central panel}: Zoom of the central region of the 
  hotspot 3C\,445 South in total intensity obtained using Briggs
  weighting. {\it Right panel}:   
  Zoom of the central region of the hotspot 3C\,445 South in total intensity
  (contours) overlaid with polarized intensity image. The first
  contour is 18 $\mu$Jy beam$^{-1}$ and corresponds to three times the
off-source noise level measured on the image plane. Contours increase
by a factor of 2. The colour scale
is shown by the wedge at the top of each image.}
\label{3C445S_FULL}
  \end{center}
\end{figure*}

\begin{figure*}
  \begin{center}
\includegraphics{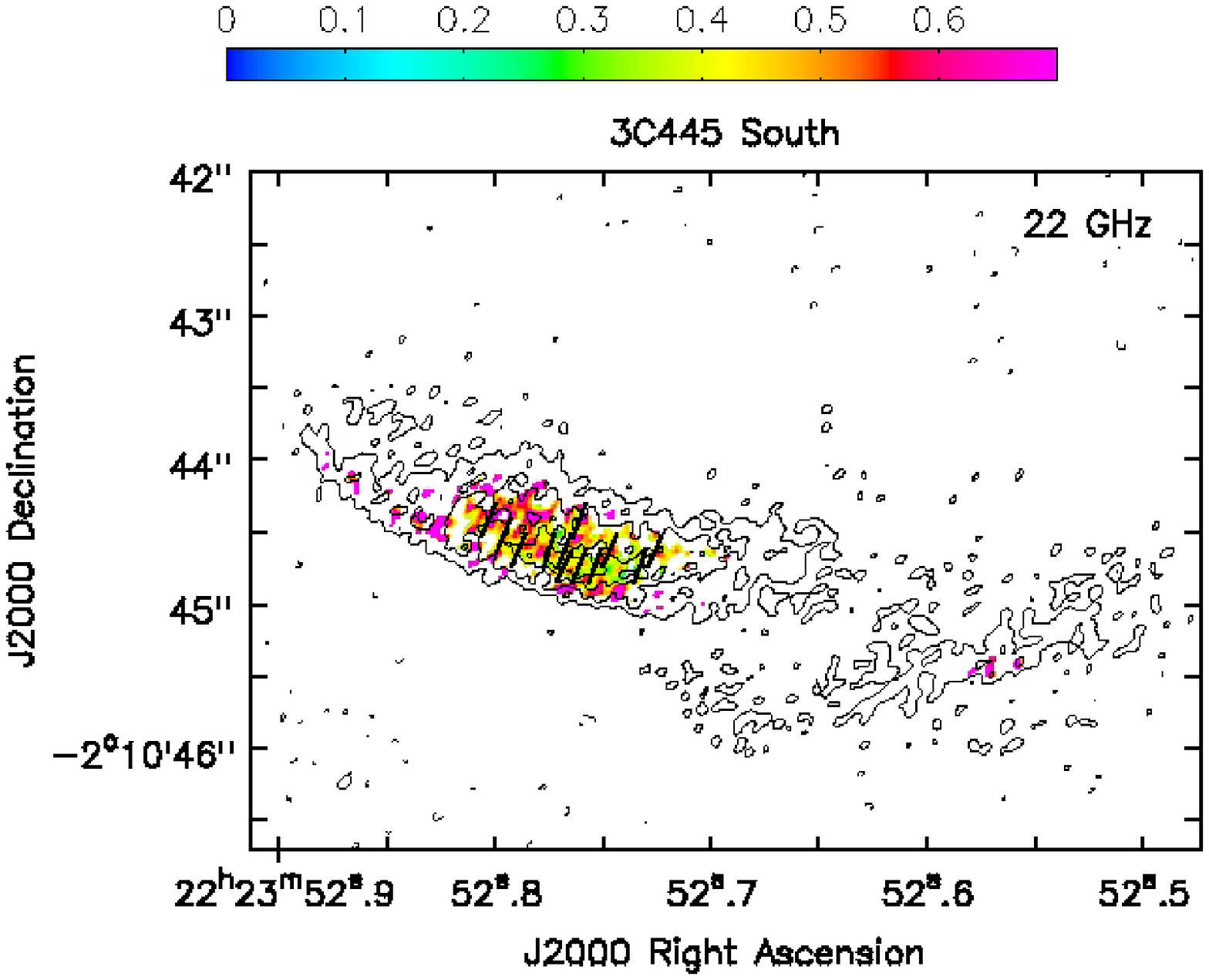}
\includegraphics{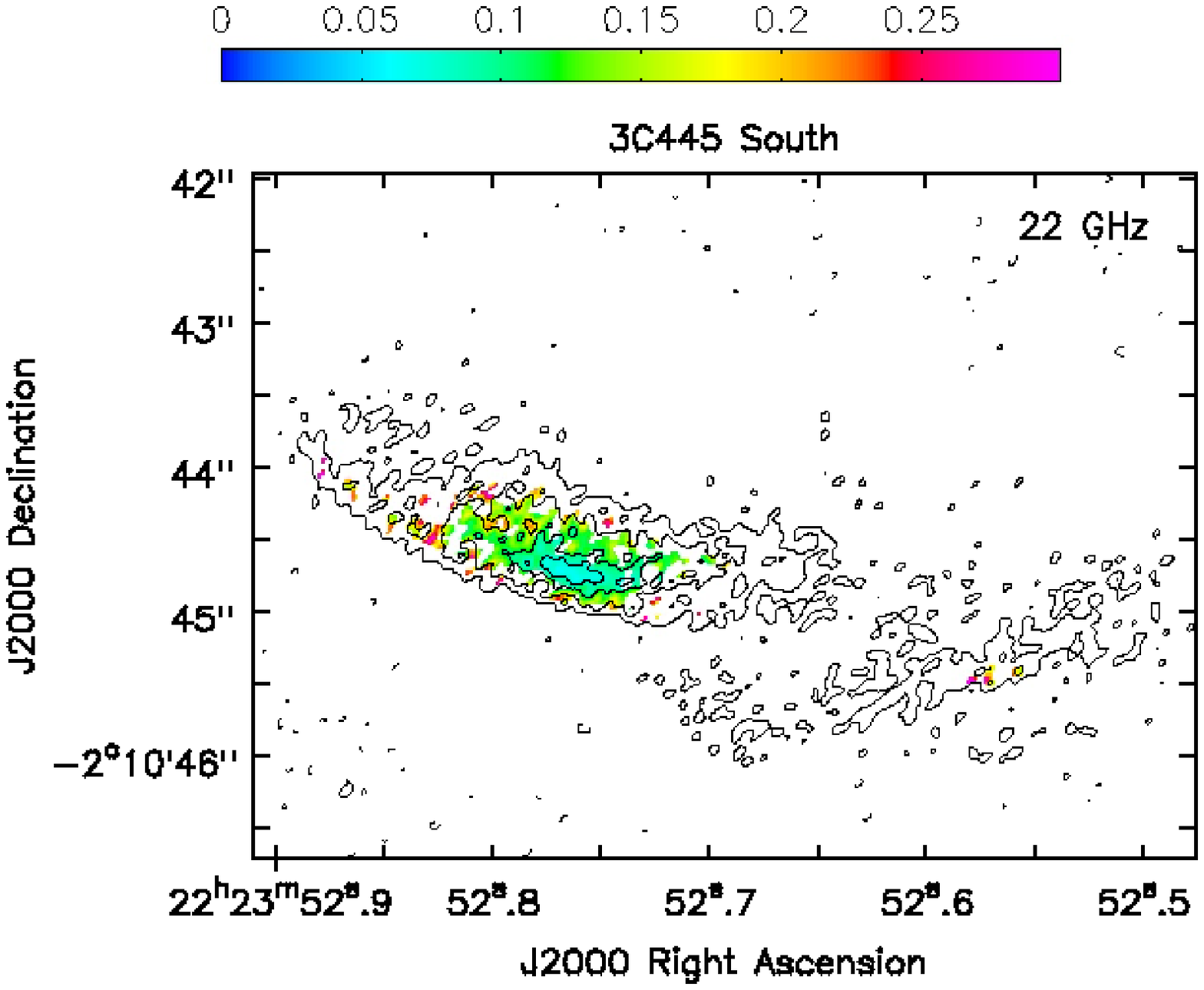} 
\vspace{8.5cm}
\caption{Fractional polarization
 image ({\it left}) and fractional
 polarization error image ({\it right}) of the hotspot 3C\,445 South.
 The first
contour is 18 $\mu$Jy beam$^{-1}$ and corresponds 
to three times the off-source noise level measured on the image
plane. Contours increase by a factor of 2. The colour scale
is shown by the wedge at the top of each image. Vectors represent the
electric vector position angle.}
\label{3C445S_POLLA}
  \end{center}
\end{figure*}

This hotspot complex has an angular size of about 6 arcsec (about 6.4
kpc) and consists of two 
main components with an 
arc-shaped structure, with the eastern one in agreement
with the edge of the X-ray emission (Fig. \ref{3c445s-chandra}). The
eastern component, labeled E in 
Fig. \ref{3C445S_FULL}, 
is the brightest one and consists of about 50 per cent of the whole hotspot
complex. Its structure is elongated in the East-West direction,
roughly transverse to the direction of the jet, as it is derived by
low-resolution images \citep[e.g.][]{leahy97,mack09}.
In this hotspot
component there are 
several unresolved polarized sub-components. The upper limit to their
angular size is between 70 and 120 mas, corresponding to about 75 and
130 pc, and a flux density of about 20$-$30 $\mu$Jy. 
The average observed fractional polarization of component E is about
36 per cent, but reaches values as high as 50 per cent. Component W is
elongated in the North West-South East 
direction and is about 
2.5 arcsec from component E in the South-West
direction. The average observed polarization of component W is 27 per cent, but
reaches values as high as 55 per cent (Fig. \ref{3C445S_POLLA}). The
fractional polarization of both components 
derived in our 22-GHz observations is slightly larger than that 
at 97.5 GHz from ALMA observations \citep{orienti17}. This may be due
to the better resolution of these VLA data ($\leq$ 0.1 arcsec) with respect
to the resolution of our earlier ALMA observations 
($\sim$ 0.5 arcsec which corresponds to $\sim$550 pc), and suggests
the presence of tangled magnetic 
field components with angular scales that are smaller than the
resolution proved by our earlier ALMA observations.\\
\indent The X-ray 
emission is detected at the edge of the main hotspot component,
likewise in 3C\,227 East. 
The polarization and the total intensity emission have
a similar structure (Fig. \ref{3C445S_POLLA}). However, the brightest
component in total intensity has a relatively low polarization and is
located about half way between the two
brightest components in polarized intensity (about 100$\pm$10 mas and
70$\pm$10 mas, respectively). 
Polarized emission is also observed in the
ridge of the arc-shaped structure and in component S that is located between
component E and W in the southern direction
(Fig. \ref{3C445S_FULL}). The EVPA changes across the whole hotspot
region, from about $-$10$^{\circ}$ to $-$55$^{\circ}$, while it
becomes roughly
perpendicular to the total intensity contours in the southern edge.
Between the two main 
components no significant polarization is detected, with a fractional
polarization below 3 per cent (Fig. \ref{3C445S_POLLA}), which is
consistent with the 4 per cent upper limit that was set by ALMA
observations \citep{orienti17}.\\

\section{Discussion}

\subsection{The hotspot structure and the magnetic field}

Particle acceleration via strong shocks, turbulence and/or magnetic
reconnection should take
place in the hotspot regions of radio galaxies. The detection of
extended synchrotron optical emission confirms the presence of
efficient particle acceleration distributed on kpc scales.  
The high resolution of our VLA observations of the hotspots in 3C\,445
and 3C\,227 pointed out a very
complicated morphology with arc-shape structures that could be
resolved into several polarized sub-components enshrouded by diffuse
emission. These structures may trace multiple shocks and compression or
regions of magnetic reconnection. The upper limit to 
their size ranges between 75 and 
160 pc. Components with size of a few tens parsecs were already observed
in other hotspots like 3C\,205 \citep{lonsdale84}, 4C\,41.07
\citep{gurvits97}, Pictor\,A 
\citep{tingay08}, and PKS\,2153$-$69 \citep{young05}. VLA
  observations of Cygnus\,A at 43 GHz could resolve the hotspot structures in
small clumps of about 0.4-kpc in size \citep{carilli99}. Depending
on their actual size, the  
existence of highly
magnetized small clumps in the hotspot regions of Cygnus\,A may
provide an explanation to the
low-energy curvature that has been observed by the Low Frequency Array
\citep{mckean16}.\\
\indent In all of the four hotspots studied here, about 10 per cent of
the total flux density of the hotspot 
region is globally produced in compact unresolved components,
whereas the majority of the
emission is on larger scales.
The polarized emission is 
characterized by several unresolved components, which may not
correspond to the unresolved components observed in total 
intensity. In particular, in 3C\,445 North the polarized
structure is different from the total intensity morphology. In this
hotspot the highly polarized components constitute about 10 per cent
of the hotspot polarized emission. The polarized emission reaches a
fractional polarization as high as 70 per cent, and is elongated
in the NE-SW direction. On the other hand, the total intensity
emission is
elongated in the NS direction 
with the southern part being brighter in total intensity, but showing
lower levels of polarization (about 20 per cent).\\

\indent The complex structure
of polarized and total intensity emission observed in these hotspots
suggests that the magnetic 
field is not homogeneously distributed in the hotspot region.
The fractional polarization between
30 and 45 per cent that is observed on average indicates the presence
of a significant random field component.
This is supported by the fact that EVPA of the
different clumps varies. The polarization decreases
at the boundaries between the clumps, and this may be due to vector
cancellation when cells with different EVPA fall within the
interferometer beam. Another piece of evidence is provided by the lower level of
fractional polarization that is observed in the images produced with
natural weights (whose restoring beam is $\sim$2.5 times larger than
the beam obtained with full resolution), where the maximum fractional
polarization is about 40 
per cent and the polarized emission is smoothly distributed across the
hotspot region. High-resolution very long baseline
interferometry (VLBI) observations are crucial for constraining the
size of the compact sub-components and infer the orientation of the
magnetic field on parsec scales.\\
\indent The observed polarization $p_{\rm obs}$ depends on the intrinsic
polarization $p_{\rm int}$ and on the number of turbulent cells $N$
intercepted by the 
observing beam $A_{\rm b}$ by
$p_{\rm obs} \sim p_{\rm int} N^{-1/2}$,
where $N = A_{\rm b} \phi l^{-3}$ in which $l$ is the size
of the turbulent cells and $\phi$ is the size of the turbulent region \citep{beck99}.
If we consider the beam of our VLA observations and the average
  angular to linear scale conversion, we have that:\\

\begin{equation}
p_{\rm int} \sim 0.1 p_{\rm obs} \phi^{1/2} l^{-3/2}   
\label{eq_turbulence}
\end{equation}

\begin{figure}
\begin{center}
\includegraphics{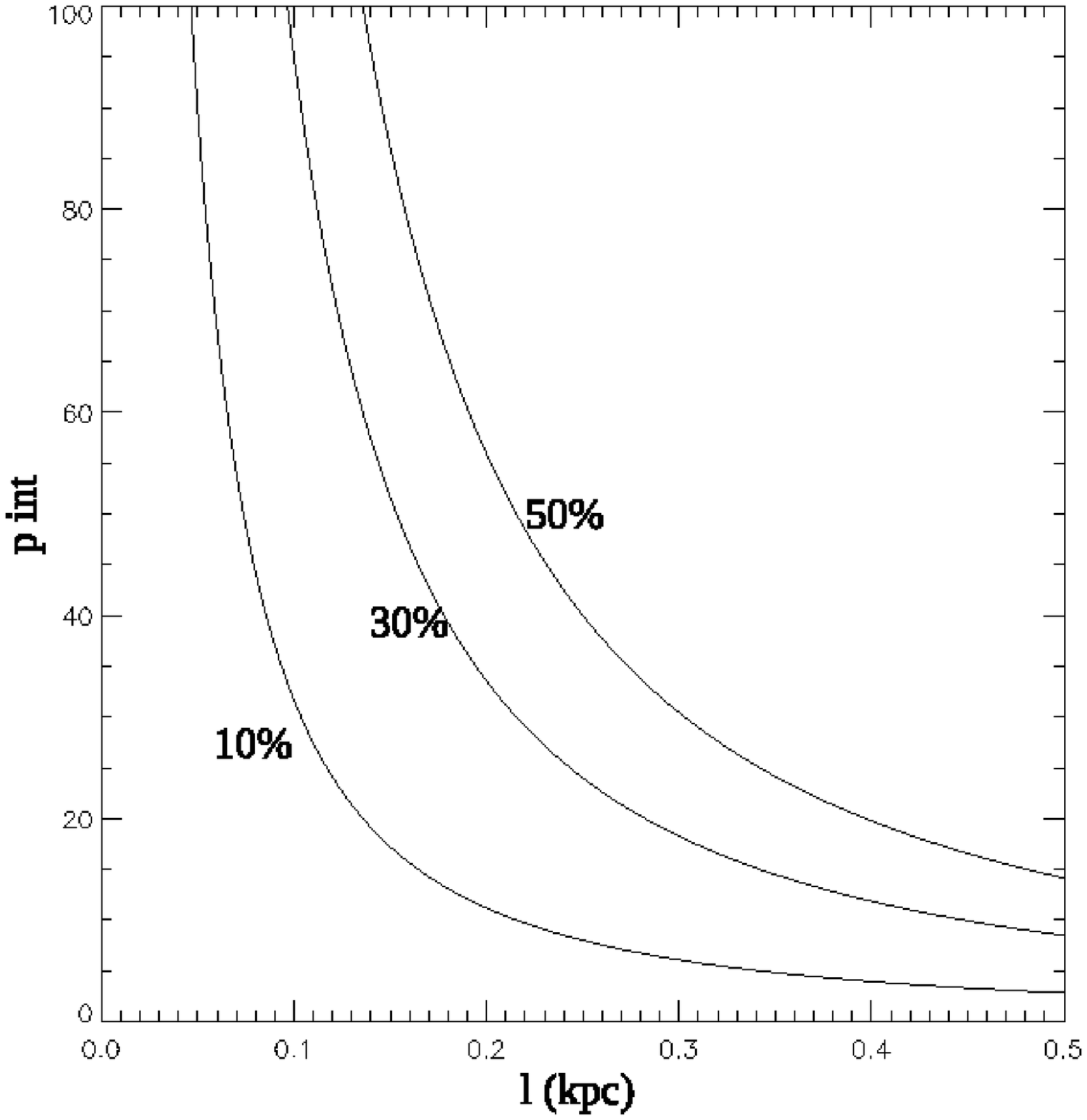}
\vspace{8cm}
\caption{Intrinsic fractional polarization as a function of the linear
  size of the turbulent cells for different values of the observed
  fractional polarization.}
\label{fig_turbulence}
\end{center}
\end{figure}

In Fig. \ref{fig_turbulence} we plot the intrinsic fractional
polarization as a function of the size of the turbulent cells,
for various values of the observed fractional polarization and
assuming a total extent
of the turbulent region of 1 kpc. This is an average value for the
main hotspot regions of the sources studied here. For $l$ of about 0.1 - 0.16 kpc, as
obtained from our VLA images and from the upper limit from ALMA
observations \citep{orienti17}, we
find that the expected intrinsic fractional polarization is between 60
and 90 per cent if we consider an observed fractional polarization of
about 30 - 45 per cent, i.e. in agreement with the average integrated
value of our observations. The high value of the expected intrinsic percentage,
above 70-80 per cent, suggests that the average scale size of the
acceleration regions and/or magnetic field length should not be much
smaller than the scale proved 
by our observations ($\sim$0.1$-$0.16 kpc).
We remark that this is an average scale size, and the presence of
transient compact pc-scale components with shorter radiative lifetime
cannot be excluded.\\

\begin{figure}
\begin{center}
\includegraphics{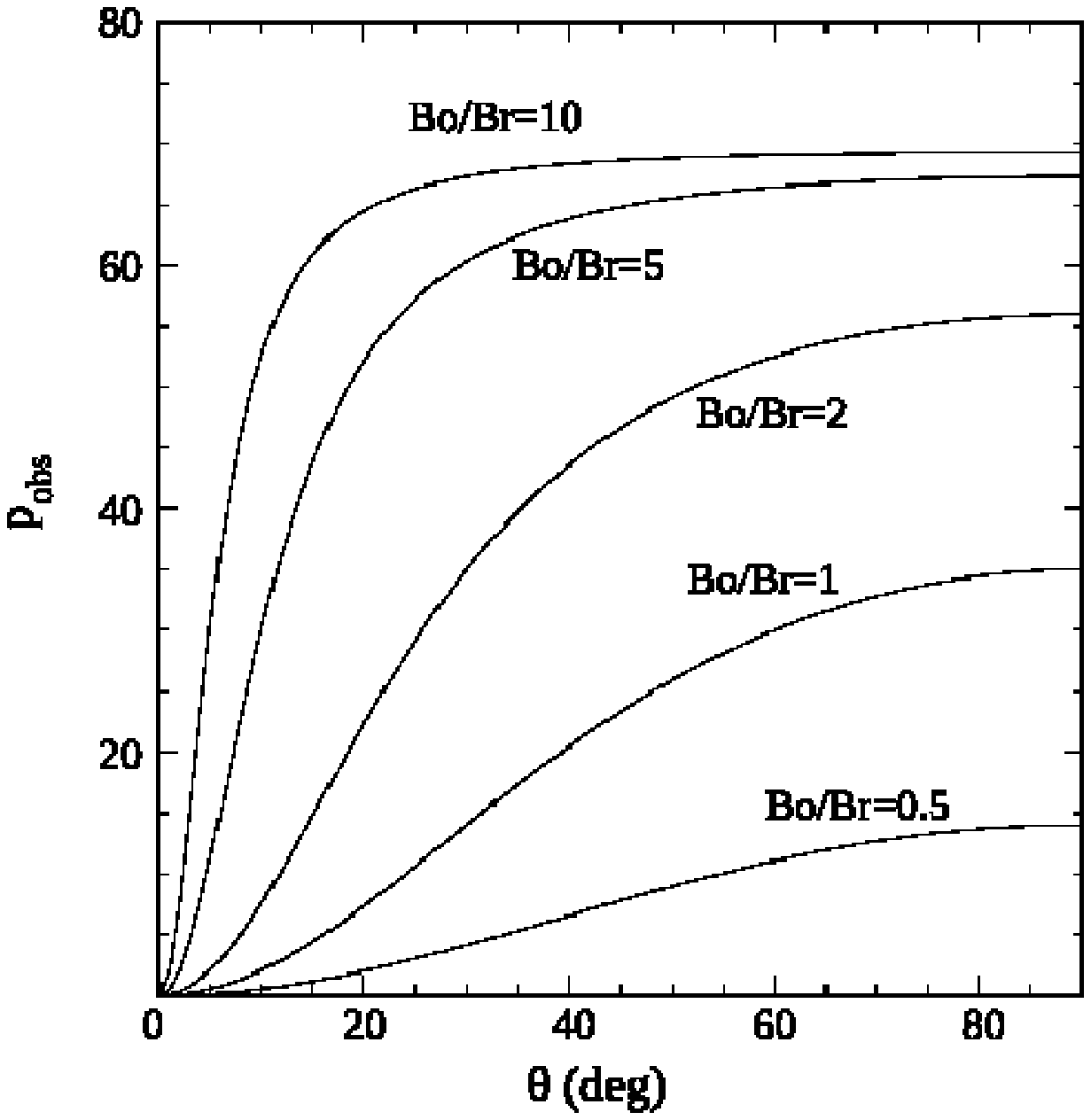}
\vspace{8cm}
\caption{Observed fractional polarization as a function of the angle
  to the line of sight for different values of the ratio $B_{o}/B_{r}$
and $p_{\rm int} = 70$ per cent.}
\label{frac_polla}
\end{center}
\end{figure}

\indent In each hotspot the EVPA is not constant and changes
arbitrarily in the central region. However, the EVPA is usually
perpendicular to the total intensity contours in the outer edge of the
hotspot structure, suggesting the presence of compression that makes
the turbulent field tangential to the boundaries of the source
\citep[e.g.][]{laing80}. \\
In presence of an irregular field that is subject to compression,
  we evaluate how the observed fractional polarization, $p_{\rm obs}$ changes as a
  function of the line of sight. The intrinsic fractional polarization
  $p_{\rm int}$ is not critically dependent on the spectral index
  $\alpha$, with a variation of $\sim$10 per cent assuming values for
  $\alpha$ between 0.5 and 1 \citep{burn66, laing80}. For simplicity,
  and for consistency with previous studies
  \citep[e.g.][]{laing80,lonsdale98}, we assume $\alpha=1$. This value
  is steeper than what is usually observed in hotspots. However, there
  seems to be a tension between the observed spectral index in
  hotspots and the injection spectral index which has been found to be
  steeper \citep{harwood17}. The observed fractional polarization,
  $p_{\rm obs}$ is:\\

\begin{equation}
p_{\rm obs} = p_{\rm int} \frac{B_{o}^{2} {\rm sin}^{2} \theta}{B_{o}^{2} {\rm sin}^{2}
  \theta + B_{r}^{2}}
\label{fractional_equation}
\end{equation}

\noindent where $B_{o}$ sin $\theta$ is the component of the ordered
magnetic field on the plane of the sky, $\theta$ is the angle to our
line of sight, and $B_{r}$ is the random
component of the magnetic field with scale length comparable to the
interferometer beam. In Fig. \ref{frac_polla} we plot the
observed fractional polarization as a function of $\theta$ for various
values of the ratio $B_{o}/B_{r}$ and assuming an intrinsic
polarization $p_{\rm int} = 70$ per cent, i.e. about the central value
in the estimated range from Eq. \ref{eq_turbulence}. The large percentage of
polarization reaching 55$-$70 per cent that is
observed in the unresolved compact components in each hotspot,
indicates that in these 
regions the magnetic field 
is organized on scales of up to a hundred pc. If the field random
component is equal to or higher 
than the ordered component, low fractional polarization can be
achieved for large angles in agreement with what we obtain from the
polarization images and from the roughly symmetric radio structure of
these two radio galaxies\footnote{For an organized
field, a low level of polarization ($p < 40$ per cent)
can be achieved only
for small angles to the line of sight, $\theta <$
10$^{\circ}$$-$15$^{\circ}$, which is unlikely for these objects.}.\\

\subsection{Multiple epochs of particle acceleration}

The two arc-shaped structures observed in 3C\,227 West
and 3C\,445 South may represent two shock fronts produced by the
change of the jet direction on large scales, creating a dentist drill
effect. On the other hand, the multiple compact components may trace
the dithering of the jet termination point(s) interacting with an
inhomogeneous back-flow material on scale size up to 100 pc. The
polarized clumps observed in the hotspots of 3C\,227 and 3C\,445  
suggest the
presence of multiple shocks that compress and align the
magnetic field. A similar result
was found for the southern hotspot in 3C\,205 \citep{lonsdale98}.
These clumps may trace the variation of the
termination point of a dithering jet, as it was suggested by
\citet{tingay08} for the hotspot of Pictor\,A, or they may represent
the highest brightness temperature regions of a wide termination shock
front with complex geometry. Determining the hotspot structure
may improve our knowledge of the high-energy emission in these regions.\\ 
\indent Compact
and transient regions may be able to produce synchrotron emission up
to X-rays. The spatial displacement between the X-ray emission and the
radio-to-optical emission challenges the synchrotron self-Compton (SSC) and the
inverse Compton (IC) scattering off Cosmic Microwave Background (CMB)
photons as the mechanisms responsible for the X-ray emission
\citep{perlman10,hardcastle07,migliori19}. IC emission
is responsible for the X-ray emission in several hotspots and jet
knots where the X-ray emission and the radio-to-optical emission are
cospatial \citep[see e.g.][]{kataoka05,stawarz07,werner12,zhang18}.
On the other hand, the
misaligned X-ray emission may be produced by compact regions of the
shock front that are  
currently accelerating 
particles, while the other shock fronts, that are traced by
radio-to-optical emission, mark the location of previous 
epochs of acceleration. 
In fact, in the presence of magnetic field strength
typical of these hotspots \citep[40$-$150 $\mu$G,][]{brunetti03,migliori19}, the
lifetime of particles emitting in X-rays is about a few hundred
years. This is marginally consistent with the average diffusive time
of the particles in regions with 100 pc in size. The expected size of
these compact and transient features emitting in 
the X-rays should be of
the order of several tens of parsecs, i.e. comparable or a bit smaller than the size proved
by these VLA 
observations. \citet{migliori19} show that the X-rays
could result from the 
integrated synchrotron emission of several compact regions,
and the radio emission of each single component is below the detection
limit at 22 GHz.\\
\indent The radiative lifetime of particles
emitting in the optical band is of the order of
10$^{4}$ yr, comparable to the average diffusive time. Optical
emission from relatively old shocks is supported by the VLT detection
of the double arc-shaped structure in 3C\,445 South and possibly in
3C\,227 West. Optical emission on kpc scale may be also produced by
turbulence, in agreement with low polarization level observed in some
regions. The highly polarized sub-components in which the kpc-scale
arc-shape structures are resolved indicates also the presence of strong
shock acceleration 
occurring in different regions and at different time. The lack of
X-ray emission cospatial with the radio-to-optical emission strongly
supports this scenario. Multi-epoch deep X-ray observations are
crucial for investigating flux 
variability and its
timescale, from months to years, which
in turn would provide us with an estimate of the size of the regions responsible
for the X-ray emission.

\section{Summary}

In this paper we presented results on sub-arcsecond polarimetric VLA
observations at 22 GHz of the hotspots of the radio galaxies 3C\,227 and
3C\,445. The conclusions we can draw from this investigations are:\\

\begin{itemize}

  \item The hotspots 3C\,227 West and 3C\,445 South are characterized
    by a double arc-shaped structure, likely indicating two shock
    fronts;\\

  \item In all the hotspots, the main components are resolved into
    several highly-polarized compact clumps, with observed fractional
    polarization that may reach values as high as 70 per cent.
    The upper limit to their size
      is about 70$-$100 mas, corresponding to about 75$-$160 pc. Their
      flux density constitutes about 10 per cent of the whole hotspot
      region, indicating that the majority of the emission is on
      larger scales;\\

  \item In all the hotspots, the highly polarised compact regions
    are embedded in the larger diffuse hotspot region, which is
    characterised by a much lower polarization, about 30 to 45 per
    cent on average, indicating significant random magnetic
    field component;\\
    
      \item Although the total intensity emission and the polarized
        emission are both patchy, their morphology may be different (e.g. 3C\,445 North). In general, 
        there seems to be a displacement between the peaks of total
        intensity and the peaks 
        of polarized emission, similar to what was already observed in
        ALMA observations of 3C\,445 South;\\

\item In 3C\,227 West and in 3C\,445 South the X-ray emission and the
  radio emission are not cospatial and the main radio component is in
  agreement with the edge of the X-ray emission. The misalignment
  is difficult to reconcile with the SSC and IC-CMB model, but it may
  be explained by a population of highly energetic synchrotron
  emitting particles. Owing to their short radiative lifetime, X-ray
  emitting electrons should mark the region where particle
  acceleration is currently taking place, while radio-to-optical
  emission is detectable in older front shocks.\\

\end{itemize}

These observations support a scenario in which shocks from a dithering
jet and turbulence are responsible for particle acceleration
up to high energies.
The highly-polarized compact regions, less than
100 pc size, resolved in these
observations may represent the local multiple particle
reaccelerators. No statistical studies on the (sub-)parsec scale
of hotspots have been performed so far, and polarimetric
high-resolution observations of a large
number of hotspots is required to assess the complex structure of
these regions and their acceleration mechanisms.

\section*{Acknowledgment}
We thank the anonymous referee for reading the manuscript carefully
and making valuable suggestions. FD acknowledges financial contribution from the agreement ASI-INAF
n. 2017-14-H.0. This work was partially supported by the Korea's
National Research Council of Science \& Technology (NST) granted by
the International joint research project (EU-16-001).
The VLA is operated by the US 
National Radio Astronomy Observatory which is a facility of the National
Science Foundation operated under cooperative agreement by Associated
Universities, Inc. This work has made use of the NASA/IPAC
Extragalactic Database NED which is operated by the JPL, Californian
Institute of Technology, under contract with the National Aeronautics
and Space Administration. This research has made use of SAOImage DS9,
developed by the Smithsonian Astrophysical Observatory (SAO). Part of
this work is based on archival data, software or on-line services
provided by ASI Science Data Center (ASDC).  
This research has made use of software
provided by the Chandra X-ray Center (CXC) in the application packages
CIAO and ChIPS.

\end{document}